\documentclass[10pt,twocolumn]{article}
\usepackage[dvips]{graphicx} 
\usepackage{subfigure} 
\usepackage{makeidx} 
\usepackage{amsmath} 
\usepackage{fancyhdr}
\usepackage{foot}
\renewcommand{\headrulewidth}{0pt}

\newcommand{\smaller}{\documentstyle[12pt]}

\newfont{\pr}{cmmib10}   \newfont{\pf}{cmmi12}   \newfont{\hi}{lasyb10}
\newfont{\ic}{cmbxsl10}  \newfont{\uu}{cmssqi8}  \newfont{\li}{cmcsc10}
\newfont{\il}{cmss10}    \newfont{\cn}{cmdunh10 scaled\magstep1}
\newfont{\kg}{cmss17}    \newfont{\kh}{cmssi17}  \newfont{\kl}{cmr17}
\newfont{\cm}{cmdunh10 scaled\magstep1}          \newfont{\ff}{cmff10}
\newfont{\su}{cmssi12}   \newfont{\mr}{cmr12}    \newfont{\ms}{cmsl12}
\makeatletter
\def\@normalsize{\@setsize\normalsize{10pt}\xpt\@xpt
\abovedisplayskip 10pt plus2pt minus5pt\belowdisplayskip \abovedisplayskip
\abovedisplayshortskip \z@ plus3pt\belowdisplayshortskip 6pt plus3pt
minus3pt\let\@listi\@listI}

\def\subsize{\@setsize\subsize{12pt}\xipt\@xipt}

\def\section{\@startsection {section}{1}{\z@}{1.0ex plus 1ex minus
 .2ex}{.2ex plus .2ex}{\large\bf}}

\def\subsection{\@startsection {subsection}{2}{\z@}{.2ex plus 1ex}
{.2ex plus .2ex}{\subsize\bf}}
\makeatother

\begin{document}
\normalsize
\date{}

\title{\Large\bf Performance of TCP/UDP under Ad Hoc IEEE802.11} 

 

\author{\begin{tabular}{c c c}
Milenko Petrovic & & Mokhtar Aboelaze\\
Dept. of ECE &  & Dept. of Computer Science\\
University of Toronto & &York University\\
Toronto ON. Canada & & Toronto ON. Canada\\
\end{tabular}}

\maketitle

\thispagestyle{empty}

\pagestyle{fancy}
\cfoot{}
\rfoot{}
\lfoot{}
\rhead{}
\chead{}
\lhead{}
\marksoff\footnotetext
{\vspace{-2mm}This research is supported by NSERC under Grant 504000. This paper was prepeared \vspace{-2mm}when the first author was at York University
\newline
\newline
\normalsize 0-7803-7661-7/03/\$17.00 \copyright 2003 IEEE}\markson

\subsection*{\centering Abstract}
{\em
TCP is the De facto standard for connection oriented transport layer protocol, while
UDP is the De facto standard for transport layer protocol, which is
used with real time traffic for audio and video.
Although there have been many attempts to measure and analyze the performance of the
TCP protocol in wireless networks, very few research was done on the UDP or the
interaction between TCP and UDP traffic over the wireless link.
In this paper, we tudy the performance of TCP and UDP over IEEE802.11
ad hoc network. We used two topologies, a string and a mesh topology. 
Our work indicates that IEEE802.11 as a ad-hoc network is
not very suitable for bulk transfer using TCP. It also indicates that it is much better
for real-time audio. Although one has to be careful here since real-time audio
does require much less bandwidth than the wireless link bandwidth. Careful and
detailed studies are needed to further clarify that issue.
}

\section{Introduction}

There is a huge growth in the wireless communication industry as can be shown
by the huge increase in the number of cellular phones, wireless LAN's
and the personal digital assistants. 
The convenience that portable computers
bring will tend to displace desktop computers. The same can be said
about wireless phones and in the future, smart appliances, which will
become commonplace. Wireless phone popularity is mainly due to freedom
of movement, that comes from the ability to use wireless phones from
virtually anywhere. Voice is the first, and still the major driving force
behind wireless technology, but the trend is to provide more services
to the user including connection to the Internet either through the
wireless phone or some other wireless device.

TCP, transfer control protocol, is the standard protocol for reliable
delivery of data over the Internet. TCP relies on IP, Internet
Protocol, for routing and data transmission. IP provides best-effort
service, which is intrinsically unreliable. This makes Internet
Protocol very simple, which is one of the reason for its popularity and
the rapid growth of the Internet. IP is the de facto standard protocol
for inter-networking. TCP is designed to go hand in hand with IP
protocol, which resulted in it becoming the dominant reliable transport
protocol.  There has been a lot of research on how to
make TCP work well in a wireline network \cite{Jacobson88a,Jacobson90a,Floyd01a}

Wireless communication is usually done in one of 2 different ways, cellular communication
or ad-hoc communication. In cellular communication, pre-established base
stations are distributed to cover the are. Each base station is responsible
of managing the mobile users in each cell. Mobile users communicate via the base
station in the cell they are in. 

The other alternative is known as ad-hoc networks.
In ad-hoc networks, there is no fixed infrastructure such as base stations or
predefined geographical cells. Mobile users are roaming
in a specified area, and they communicate by sending (receiving) messages to (from)
each other. If two users are close enough to each other they can communicate
directly. If the users are far apart then the rest of the users can forward packets
to and from these two users in order to be able to communicate. That means every 
mobile user serves as a relay or a router in order for all the nodes to be able
to communicate. Several routing protocol were proposed for ad hoc networks
\cite{Johnson96a, Park97a, Perkins97a,
Perkins94a, Pei00a, Pei00b}

Wireless medium is a difficult medium for communication.
In free space, a typical wireless channel is susceptible to the problems
of path loss, shadowing, multipath fading and interference. Usually the bit rate 
error for wireless channels is higher than wireleine channels,
and its bandwidth is less. that makes using the wireless link with a protocol that
was specifically designed for a wireline networks a bit challenging.

There has been a lot of research trying to measure and analyze the performance of
TCP over wireless links for both cellular, and ad-hoc networks. However,
for applications like audio or video, usually that will be carried out
using UDP instead of TCP. Very few studies were carried out on the performance
of UDP on wireless links, or in the interaction between TCP and UDP traffic over
wireless links. In this paper, we present simulation results of
the interaction between TCP and UDP traffic (both real-time and bulk)
over ad-hoc networks
using IEEE802.11 as a wireless link \cite{standard97a}.

The organization of the paper is as follows, In section 2 we review some
of the previous attempts in measuring TCP performance over wirless networks.
In section 3, we present our network setup and the error model we will use throughout
experiement. Section 4 presents some results on a string of wirless nodes, while
section 5 presents results for a mesh topology. Section 6 is a conclusion.

\section{Previous Work}

There is a large volume of literature on the performance of TCP in wireless environment.
Research on improving the performance
of TCP over wireless networks can be classified into two categories.
improvements at the link layer and
improvements by making modifications
to TCP. We will very briefly mention some of the previous work and classify it.

Snoop protocol \cite{Balakrishnan95a}
is designed to be TCP aware, and to mask unreliability of wireless
layer. Snoop is implemented as a layer in TCP/IP architecture stack. It
is located just below TCP layer. Snoop can be located at both the access
point and the mobile nodes. It is not necessary to use it at mobile
nodes, which makes it easier to implement, but transfer of data from
mobile host to wired node will not benefit from snoop. Snoop at
the access point is only able to improve TCP performance of connections
from wired host to mobile hosts.

Explicit Feedback
(EF) \cite{Bakshi97a} is a mechanism used by the access point to inform
TCP sender (located in the wired network) that wireless channel is
currently experiencing a lot of errors and that it should not invoke
congestion avoidance procedure on lost segment timeouts. This requires
modifications at both the access point and the TCP sender. The explicit
feedback messages are sent to the sender after every failed transmission
to a mobile node from the access point.
In \cite{Cobb95a} access
point is assumed to send acknowledgments to senders on the wired network
for every segment it receives. These acknowledgments indicate to the TCP
sender the segment reached the access point and if it does not receive the
acknowledgment for it, then the sender can assume that the loss occurred
due to corruption over wireless medium, and congestion avoidance should
not be initiated.

The last hop acknowledgment scheme assumes that losses
over wireless network happen only due to corruption and that wireless
network is the last hop on the TCP segment path (which is the case for
cellular networks).  The acknowledgment from the access point is called
last hop ACK (LHACK). In the case that TCP sender does not receive LHACK,
then congestion in the wired network caused packet to be dropped and
therefore TCP sender should start  congestion avoidance procedure.

In \cite{Biaz99a} , TCP segment
inter-arrival times at TCP receiver are used to distinguish between
congestion and wireless losses. It is assumed that TCP segments will
queue at the access point in the case when TCP receiver is on a wireless
node. Queuing occurs here because of small wireless bandwidth as compared
to wired bandwidth. TCP receiver looks at inter-arrival time between every
segment. If the inter-arrival time between two segments is a multiple of
a segment transmission time over wireless network, but the two segments
arrived out-of-order, then TCP receiver assumes that all segments between
last in-order received segment and the segment just received are lost
due to congestion in the wired network. This scheme assumes that due
to queuing at the access point, all segments will be sent back-to-back
to the wireless node. It also assumes that there is no congestion in
the wireless link and that only bulk transfers are used. In the case
that segments are lost because of congestion, the queue at the access
point will have gaps in sequence numbers, but inter-arrival times at
the mobile node will be the same for all packets. From these gaps, TCP
receiver can conclude that congestion is the cause of the losses. On the
other hand, if losses occurred in wireless part then the inter-arrival
times will not be a multiple of segment transmission times.  From this,
TCP receiver can conclude that the losses occurred because of wireless
error and it does not initiate congestion avoidance.

Mobile-TCP \cite{Stangel98a}
is another solution that is designed mostly for problems of
disconnections. Mobile-TCP informs TCP-sender (on wired network)
that a disconnection occurred. If TCP sender detects a loss (duplicate
acknowledgments or timeout) it will perform retransmissions but without
reducing its send window. Once disconnection ends, TCP sender is informed
to resume normal operation.

\section{Experiment Setup}

In this paper, a series of simulations is performed to determine
the perofrmance and the interaction between TCP carrying bulk traffic
and UDP carrying real-time audio traffic in wireless links.
We used the ns-2 simulator \cite{Breslau00a} with the wireless extension form the Monarch
project at CMU \cite{Monarch99}.
The main perofrmance criteria for bulk transfer is the throughput. While
the main performance criteria for real-time audio is cell loss ration.
These sets of simulations are similar.
The cell loss in UDP traffic is mainly due to two different factors.
A cell (frame) is lost if it will be tranmsitted up to the maximum number
of times and always is delivered in error. Or if the cell is delayed due to
queueing or multiple transmission up to the maximum delay limit, in this case
it is not useful anymore and will not be transmitted and is dropped by the sender.
to those in \cite{Gerla99a}, although they only used bulk transfer with TCP
without FEC. In all the setups DSDV is used as the
routing protocol.

The model of errors in a wireless channel is Gilbert-Elliot
\cite{Gilbert60a,Elliot63a}, which captures bursty nature of errors
in radio channels. It is a time-based two state Markov chain, where a
``good'' state has a low error probability ($10^{-6}$) and a ``bad'' state
has a high error probability ($10^{-2}$) (same as in \cite{Kim00a}).
The average length of the states is exponentially distributed with
mean duration for ``good'' state of 0.1 second and 0.0333 seconds for
the ``bad'' state. In two-sate Markov chain, ``good'' state is always
followed by a ``bad'' state and vice versa. Each node uses the error
model independently, which means that each nodes sees the radio channel
differently.  The original error model for wireless channel in ns2 has
been modified to correct an error in its operation.  The model failed
to make any state transitions when the channel was idle regardless of
the passage of time. The consequence of this is that states lasted for
very long time.

In order to deal with errors, we used Reed-Solomon FEC in order to detect and
errors when the channel is in the bad state. The choice of the code was such
that the channel will have the same BER in the good state and the bad setate with FEC.
That results in decresing the efficiency of the TCP by 40\% due to the overhead
of the FEC. Thus we eliminated the bad state on the expense of a reduced 
bandwidth.

We used two different topologies, first we used a linear string of 8 nodes
where every node can communicate with its
two neighbors only (one neighbor in case of the end nodes). Then we used a mesh
topology where every node can
communicate with its four neighbors in the row and column directions.
For a complete results description, the reader is referred to \cite{Milenko02}.

\section{String}
\index{Evaluation!string}
\label{sec:string}
\subsection{Single bulk TCP transfer}

Using a string topology we examine the
performance of a multi-hop TCP connection.  In this configuration,
every node is only able to communicate with its immediate neighbor, so
routing is needed to reach nodes that are not within transmission range.
The source node initiates a bulk TCP transfer to one of
the other nodes. The measure of performance is throughput. All nodes are
assumed to be stationary so routing has no effect on the throughput, thus
TCP performance depends mainly on MAC protocol performance. We look at TCP
throughput for connections between nodes 0-1, 0-2, 0-3, 0-4, 0-5 and 0-6.
we ran the simulation using FEC, without FEC, and without any errors (ideal channel)
for comparison.

\begin{figure}
     \centering
     \subfigure[]
     {
          \label{fig:string-norts-200}
          \includegraphics[width=.22\textwidth,height=0.4\textwidth,angle=-90]
          {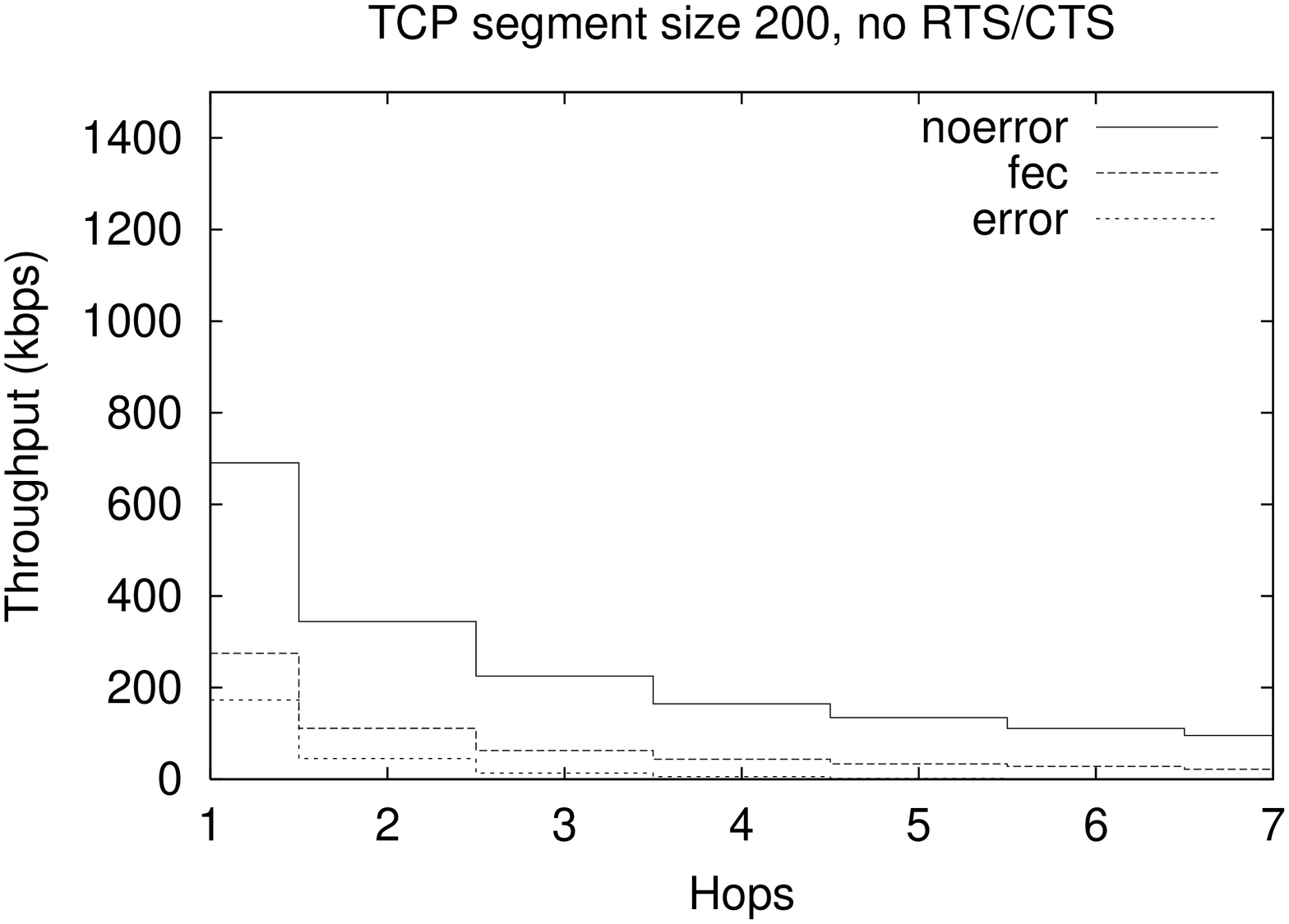}
     }
     \subfigure[]
     {
          \label{fig:string-rts-200}
          \includegraphics[width=.22\textwidth,height=0.4\textwidth,angle=-90]
          {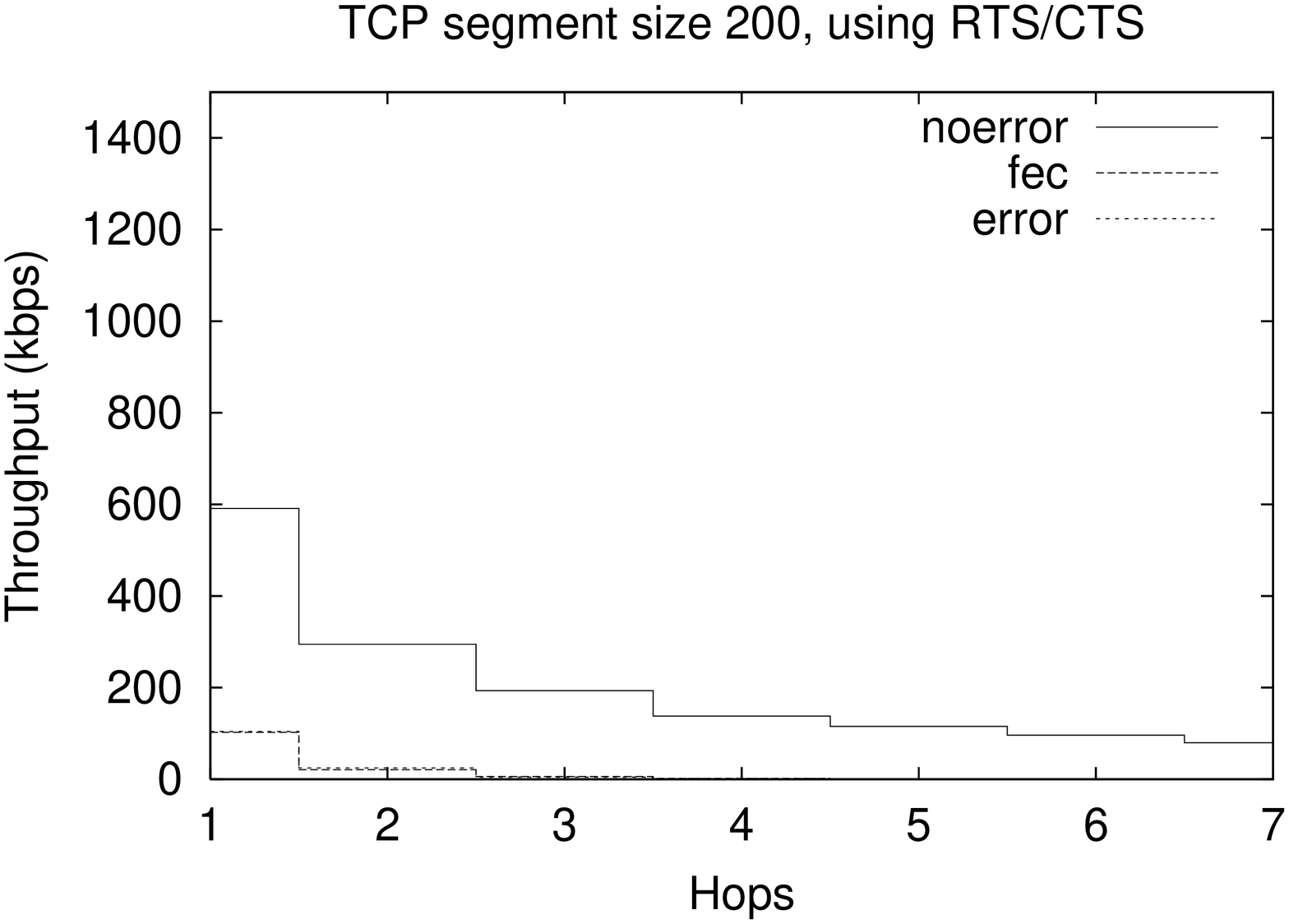}
     }
     \subfigure[]
     {
           \label{fig:string-norts-1000}
           \includegraphics[width=.22\textwidth,height=0.4\textwidth,angle=-90]
           {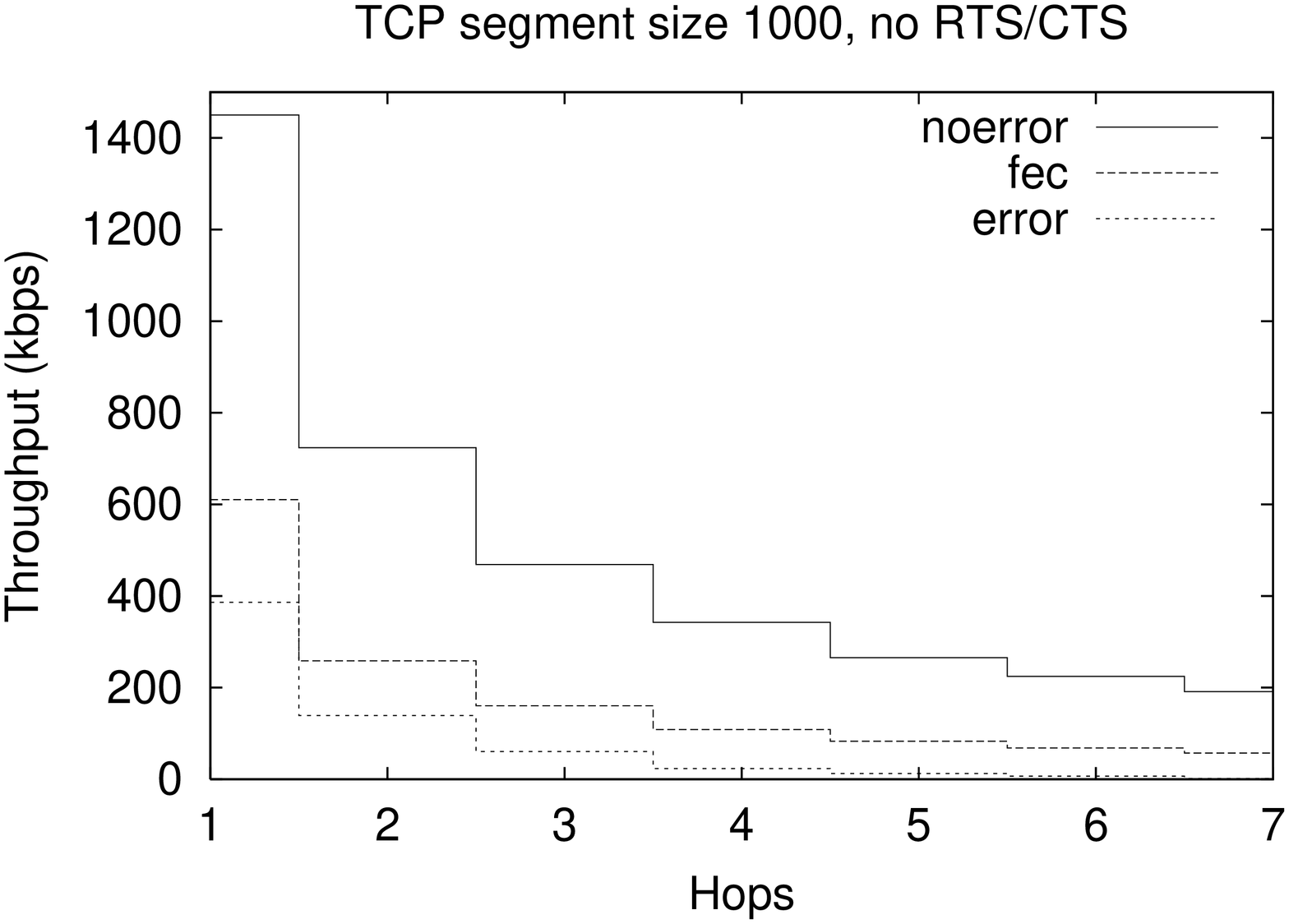}
     }
     \subfigure[]
     {
           \label{fig:string-rts-1000}
           \includegraphics[width=.22\textwidth,height=0.4\textwidth,angle=-90]
           {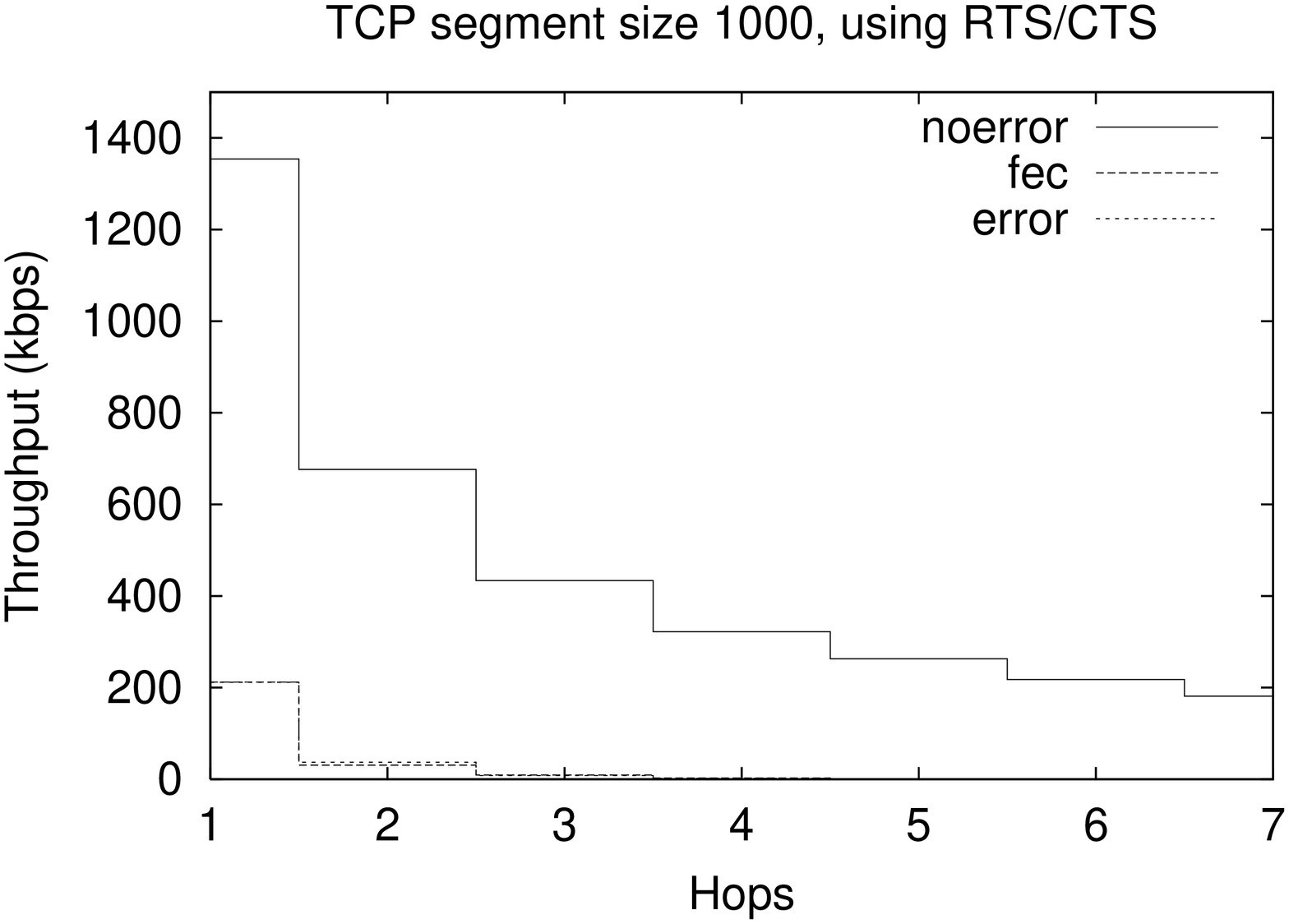}
     }
     \caption[String - TCP throughput]{String: TCP window size = 1}
     \label{fig:string}
\end{figure}

In figures \ref{fig:string} we notice that a larger segment size produces better reults
than a smaller one. Also, RTS/CTS is almost having a negative impact on the
peformance, and the system performs better without collision avoidance.

\begin{figure}
     \centering
     \subfigure[]
     {
          \label{fig:string-200-1}
          \includegraphics[width=.22\textwidth,height=0.4\textwidth,angle=-90]
          {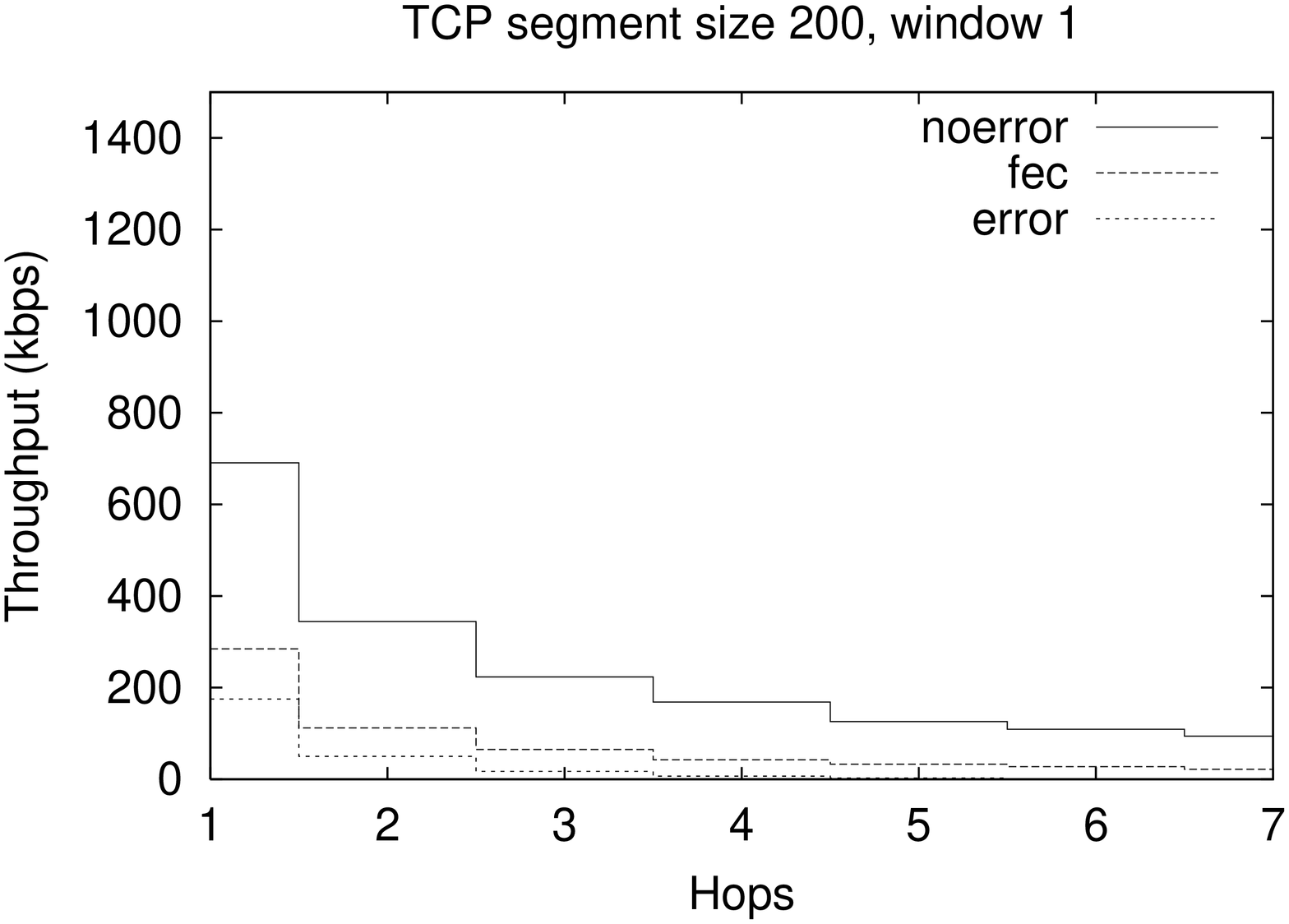}
     }
     \subfigure[]
     {
           \label{fig:string-200-4}
           \includegraphics[width=.22\textwidth,height=0.4\textwidth,angle=-90]
           {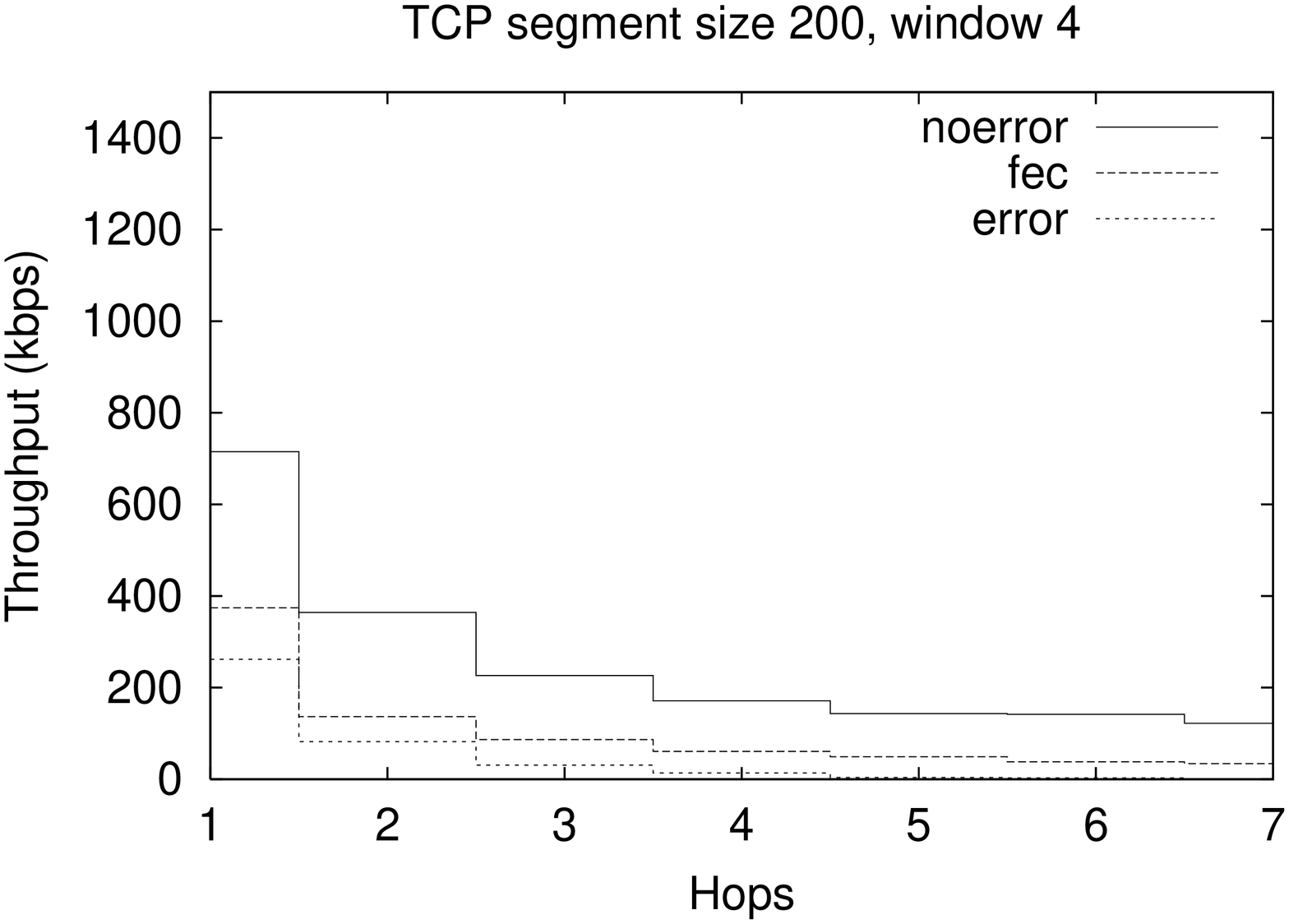}
     }
     \subfigure[]
     {
           \label{fig:string-200-16}
           \includegraphics[width=.22\textwidth,height=0.4\textwidth,angle=-90]
           {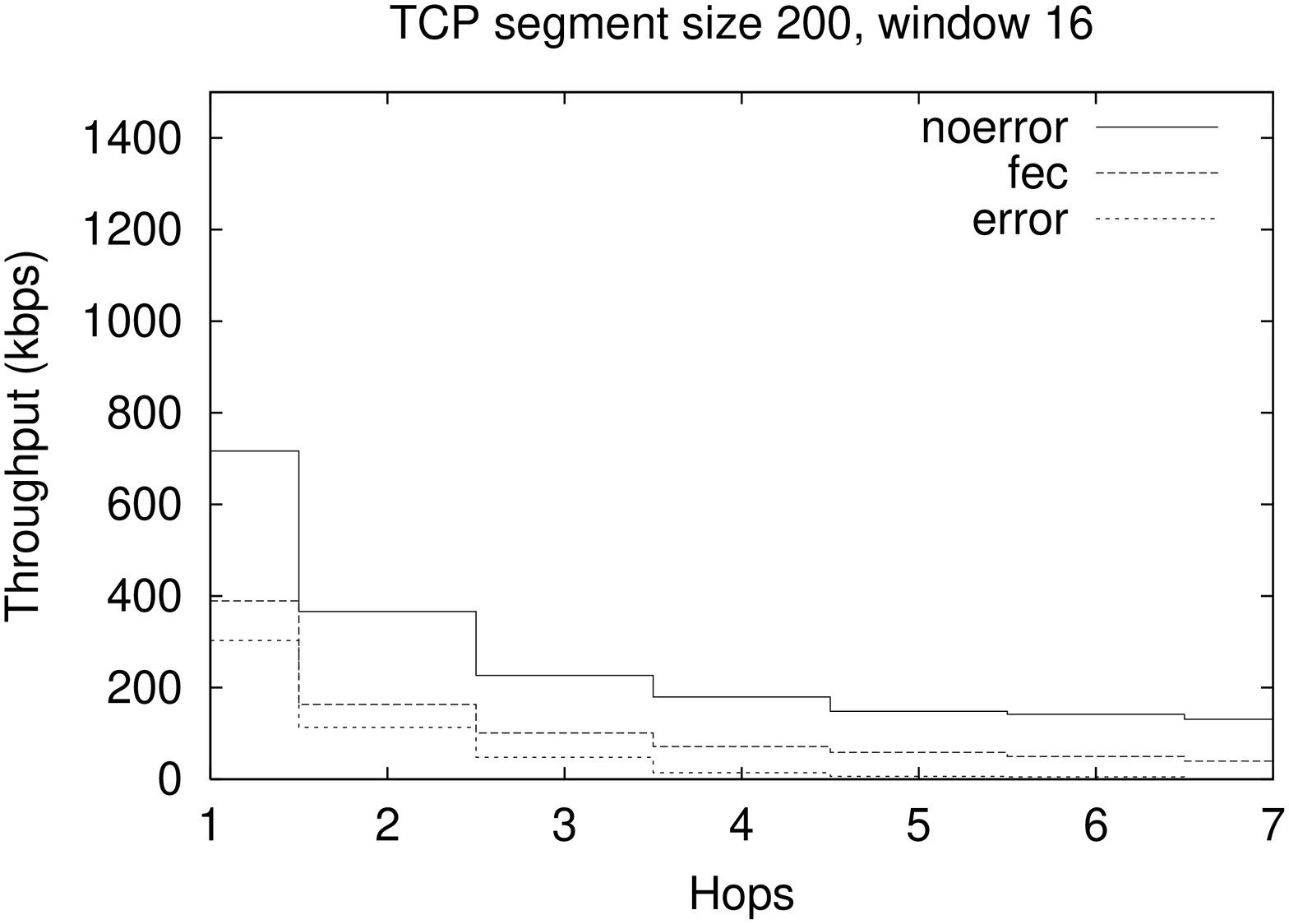}
     }
     \subfigure[]
     {
           \label{fig:string-200-32}
           \includegraphics[width=.22\textwidth,height=0.4\textwidth,angle=-90]
           {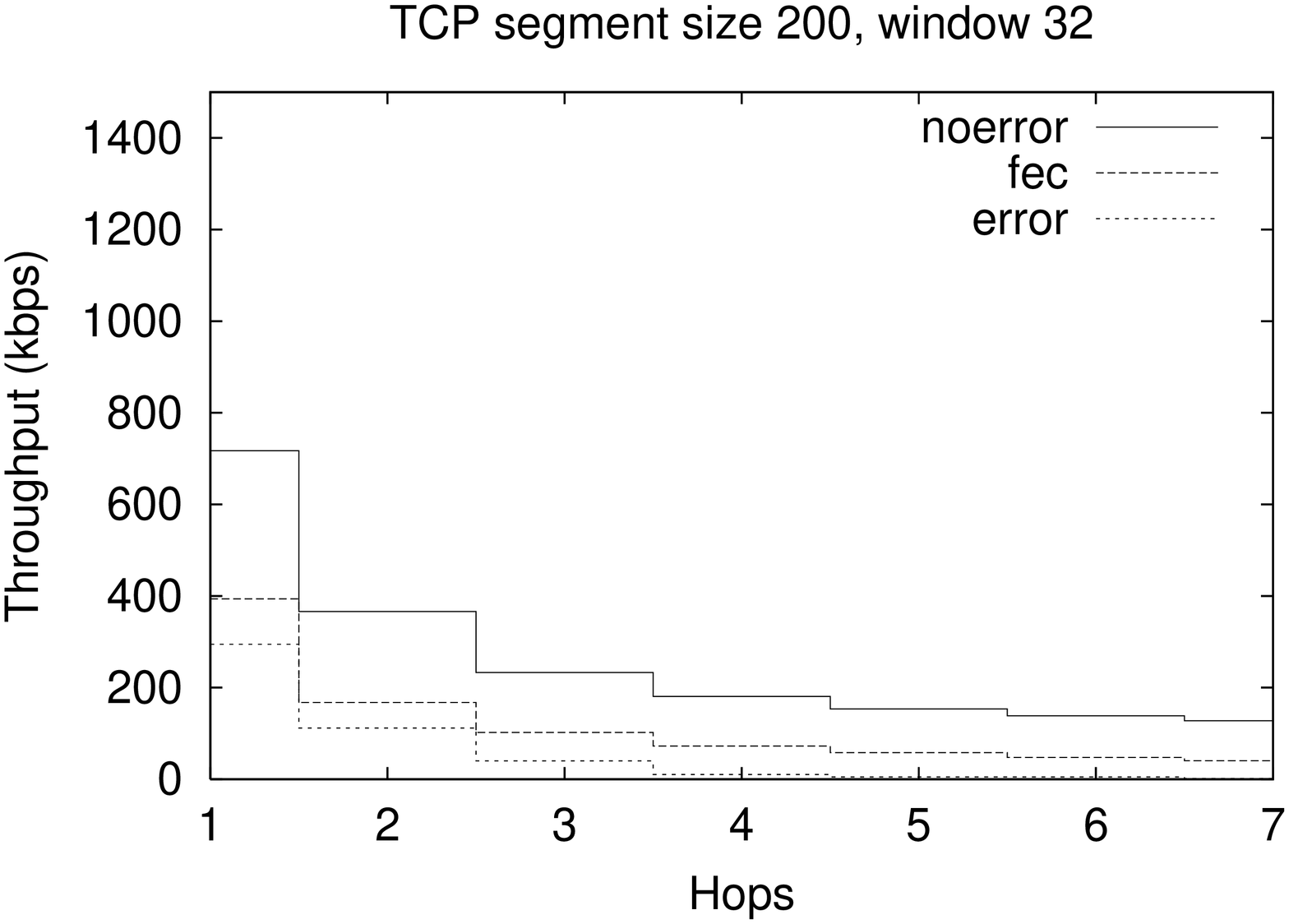}
     }
     \caption[String with small packets]{String: TCP segment size = 200 bytes}
     \label{fig:string-200-window}
\end{figure}

\begin{figure}
     \centering
     \subfigure[]
     {
          \label{fig:string-1000-1}
          \includegraphics[width=.22\textwidth,height=0.4\textwidth,angle=-90]
          {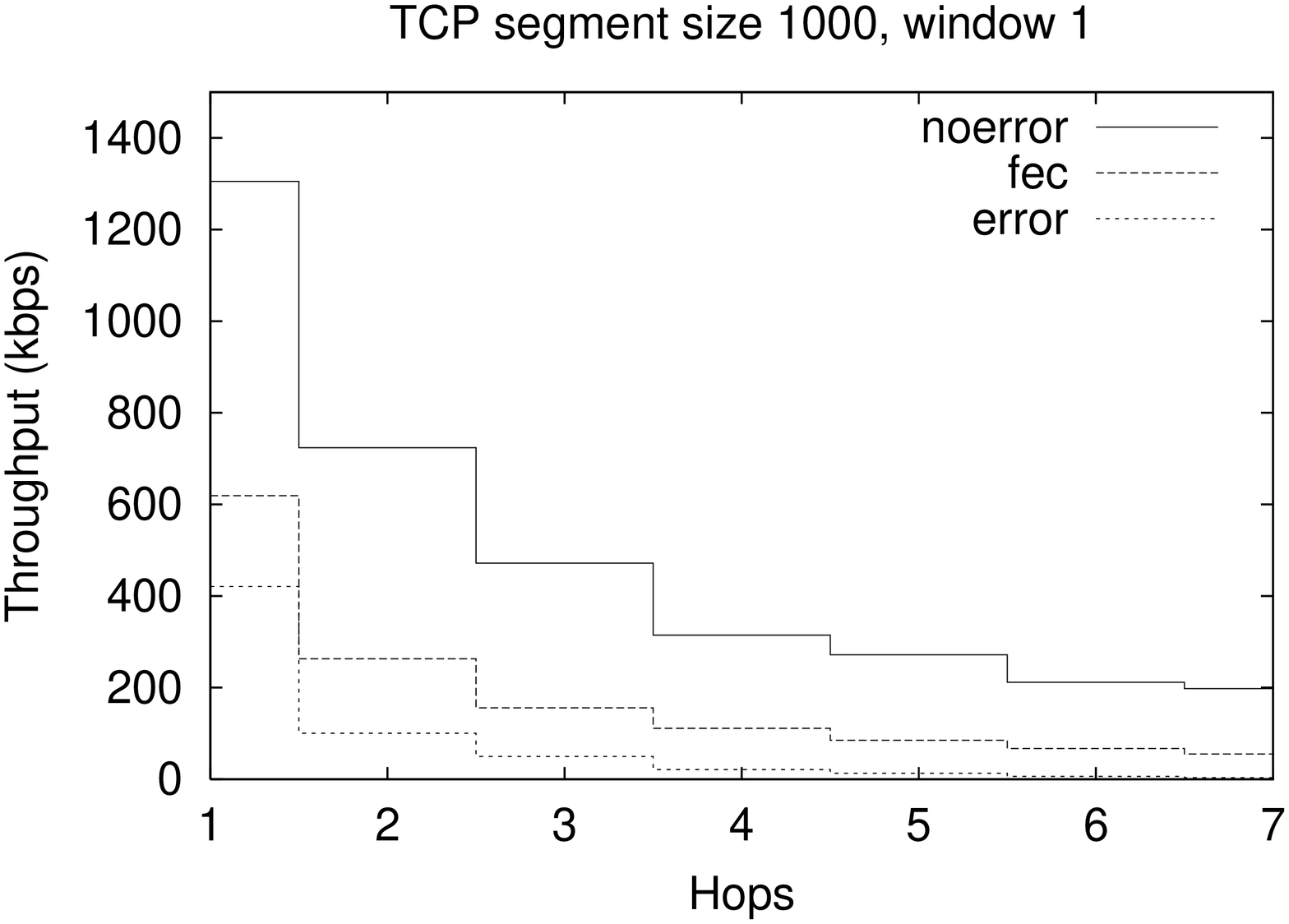}
     }
     \subfigure[]
     {
           \label{fig:string-1000-4}
           \includegraphics[width=.22\textwidth,height=0.4\textwidth,angle=-90]
           {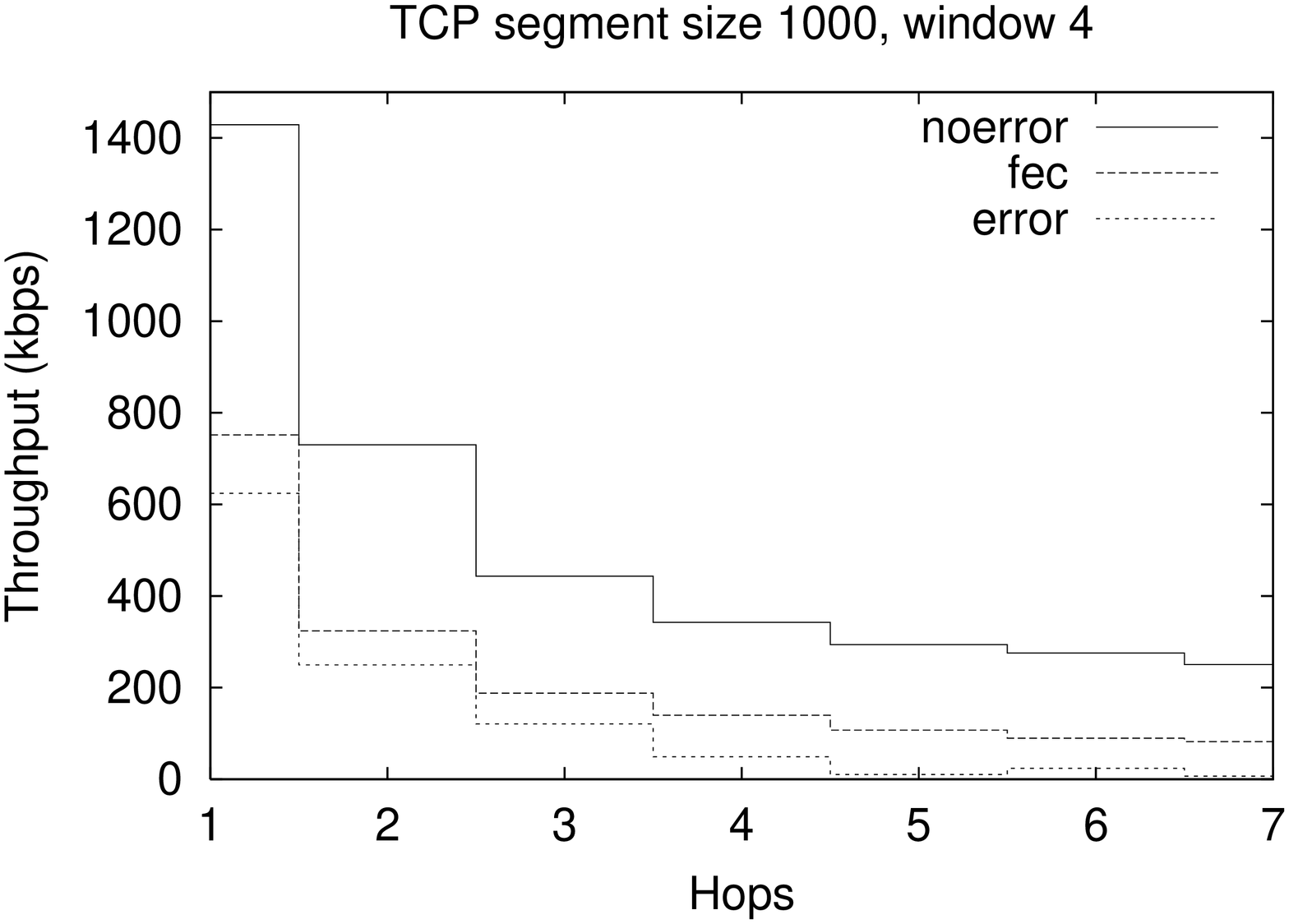}
     }
     \subfigure[]
     {
           \label{fig:string-1000-16}
           \includegraphics[width=.22\textwidth,height=0.4\textwidth,angle=-90]
           {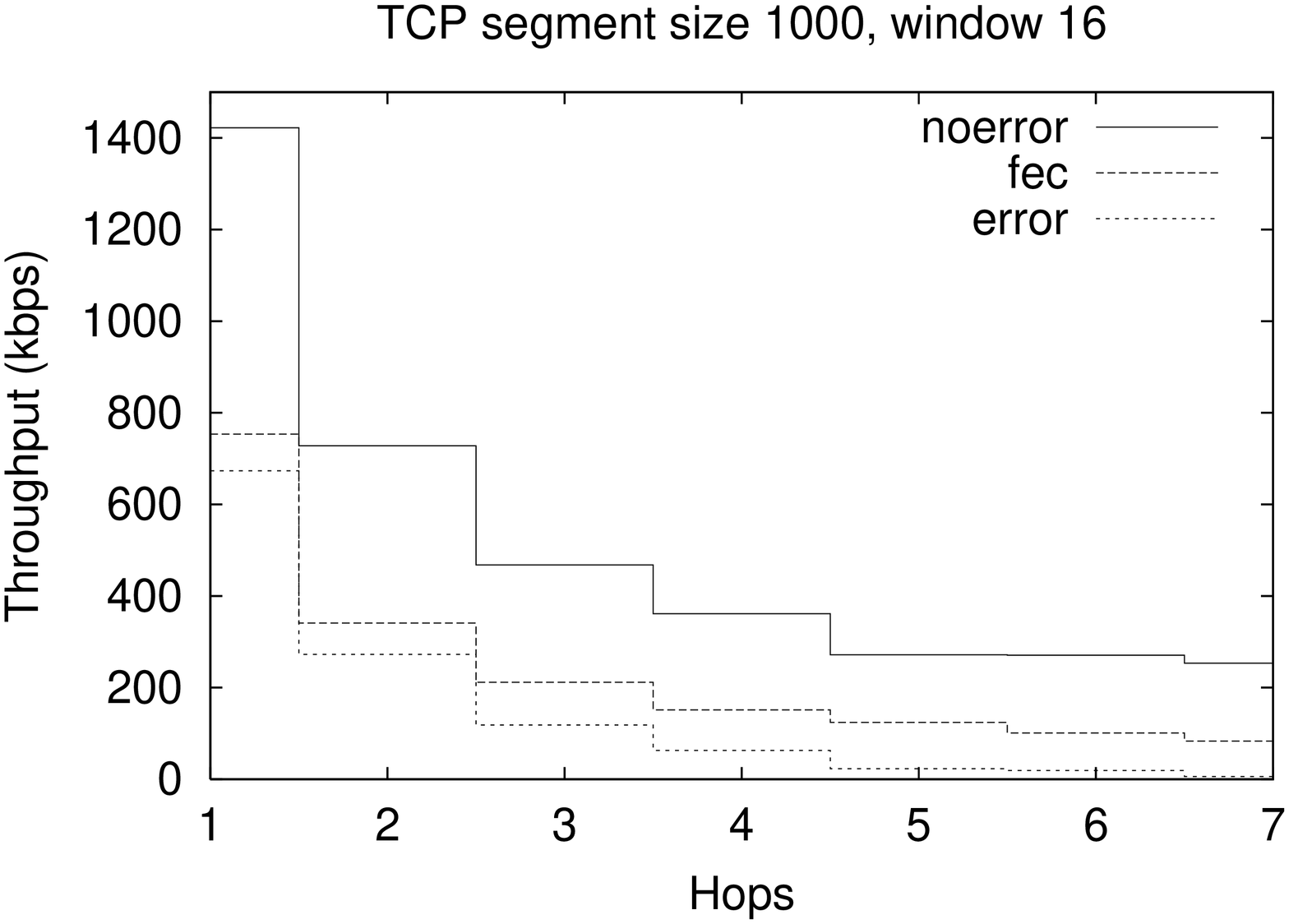}
     }
     \subfigure[]
     {
           \label{fig:string-1000-32}
           \includegraphics[width=.22\textwidth,height=0.4\textwidth,angle=-90]
           {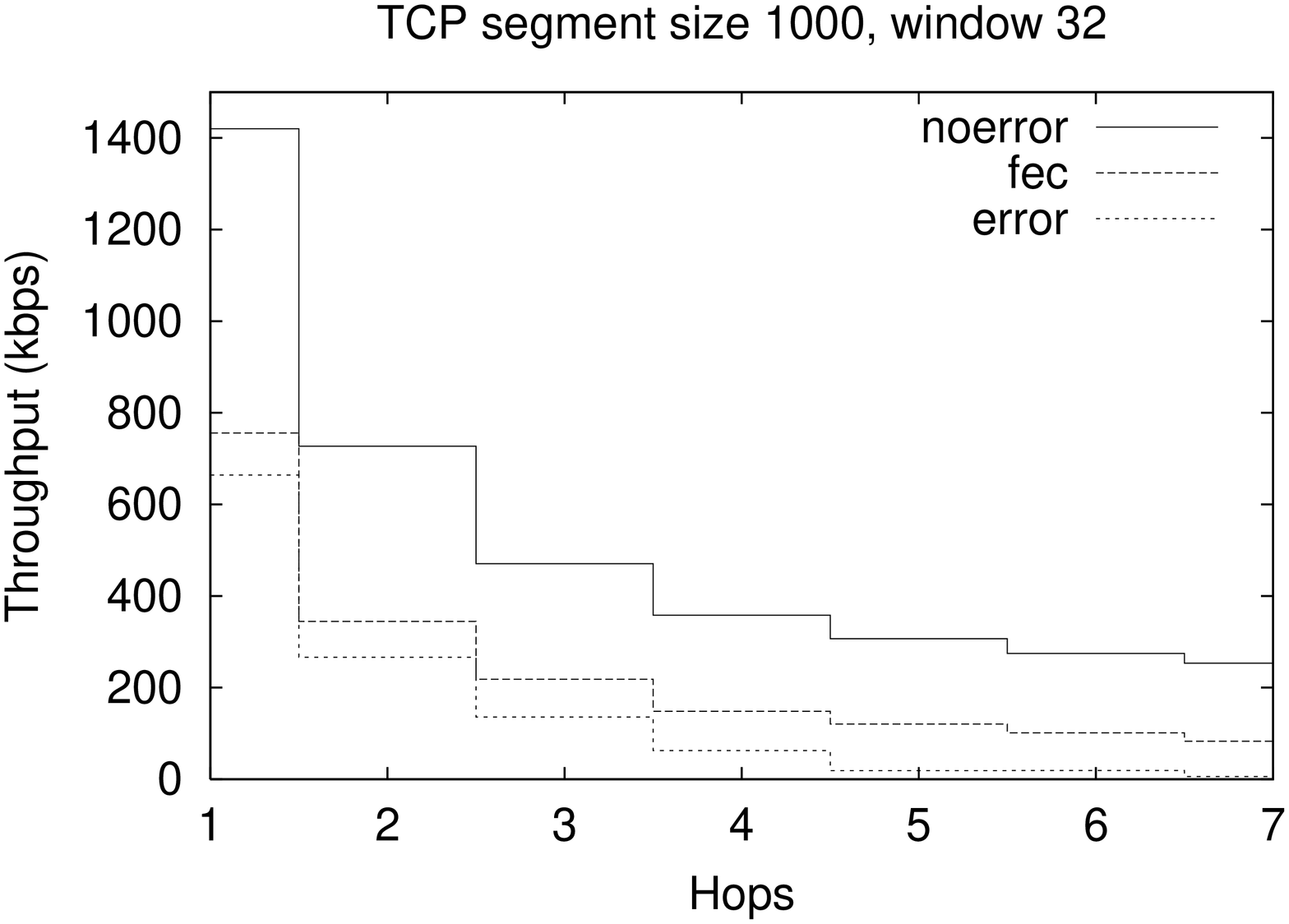}
     }
     \caption[String with large packets]{String: TCP segment size = 1000 bytes}
     \label{fig:string-1000-window}
\end{figure}

In figures \ref{fig:string-200-window} and \ref{fig:string-1000-window}
it can be seen that as number of hops increases, the use of larger
window size results in increase in throughput. With large window size,
TCP can have more segments to transmit at each node without waiting to receive the ACK
for the transmitted segments (many will be lost and the TCP will
perform slow start).  It is therefore recommended to allow TCP to use larger
window size at all times. Furthermore in figure \ref{fig:string}, we can
see that with errors, plain TCP connections are barely able to transfer
packets more then 4 hops away. With increased window and large packet
size as in figures \ref{fig:string-1000-16} and \ref{fig:string-1000-32},
plain TCP is able to complete transfers, but at a very low throughput.

\subsection {Audio}

Here, we consider combination of real-time audio and bulk transfer.
The objective is to investigate the interaction between real-time audio using
UDP, and bulk transfer using TCP.
First, we consider audio only, we simulate and measure the loss percentage
for a call from node 0 to the seven other nodes.
Table \ref{tab:voice+bulk} shows the loss percentage for a call from
node 0 to nodes 1..7. As expected the loss ratio increases with the number of hops
(even for a single call), and more than few hops results in increasing the loss rate 
beyond the generally accepted 1-2\%. We also notice that smaller UDP packet size
leads to a better loss rate. We believe that although a smaller UDP packet
means more packets, however it also means that by not waiting to collect
a large packets, we can tranmsit cells with minimum delay thus reducing the
probability of time out and discarding the cell in case of multiple retramsissions.

\begin{table}
\centering
\begin{tabular}{|c|c|c|c|} \hline
 &\multicolumn{3}{c|}{UDP Packet Size} \\ \hline
>From node 0 to & 600 bytes &400 bytes  & 300 bytes  \\ \hline
1 & 0.25\% & 0.3\% & 0.4\% \\ \hline
2  & 0.31\%  & 0.85\% & 0.75\% \\ \hline
3  & 1.5\% & 1.3\% & 1.2\% \\ \hline
4  & 2.05\% & 1.4\% & 1.7\% \\ \hline
5  & 2.1\% & 2.0\% & 1.7\% \\ \hline
6  & 2.6\% & 3.5\% & 4 \% \\ \hline
7  & 6.0\% & 6.0\% & 4.0\% \\ \hline
\end{tabular}
\caption[bulk and voice string]{loss ratio for a single real-time audio call from node 0 to the rest of the nodes}
\label{tab:voice+bulk}
\end{table}

Next, we run two experiments with multiple voice connections.
The first experiment, we run two way audio connection between nodes 0-7, 1-6, 2-5,
and 3-4. In the second, we run 0-1, 2-3. 4-5, and 6-7. The first configuration
produces the maximum overlaps between these four connections, while the later
produce the minimum overlap. Tables

\begin{table}
\centering
UDP Packet Size
\begin{tabular}{|c|c|c|c|c|} \hline
100 & 200 & 300 & 400 & 600 \\ \hline
19.7\% & 2.27\% & 3.23\% & 3.54\% & 3.13\% \\ \hline
\end{tabular}
\caption[stringv1]{Average loss ratio for 4 overlapping audio connections}
\label{tab:stringv1}
\end{table}

\begin{table}
\centering
UDP Packet Size
\begin{tabular}{|c|c|c|c|c|} \hline
100 & 200 & 300 & 400 & 600 \\ \hline
0.36\% & 0.42\% & 0.36\% & 0.33\% & 0.47\% \\ \hline
\end{tabular}
\caption[ stringv2]{Average loss ratio for 4 non-overlapping audio connections}
\label{tab:stringv2}
\end{table}

Table \ref{tab:stringv1} shows the loss ratio for the first configuration, while
Table \ref{tab:stringv2} shows the loss ratio for the second configuration. 
We notice that for multiple audio connections a UDP packet size of 300 bytes produces
the best results.

Table \ref{tab:vtp07} shows both the throughput of one TCP connection and one
audio connection between nodes 0 and 7. We notice that with a large packet size the TCP
throughput is 0. We also notice that although a larger window size increases
the TCP throughput, it also increases the loss ratio for UDP packets.

\begin{table}
\centering
\begin{tabular}{|c|c|c|c|} \hline
 &\multicolumn{3}{c|}{Window Size in packets} \\ \hline
Packet Size & 1 & 2 & 4 \\ \hline
500 & (32, 5.0\%) & (37, 4.6 \%) & (34, 14\%) \\ \hline
1K & (46, 5.0\% & (54, 5.0\%) & (66, 6.6\%) \\ \hline
2K & (4.8, 5.0\% & (8.3, 8.5 \%) & (7, 4.5\%) \\ \hline
4K & (0.0, 4.0\%) & (0.23, 5.5\%) & (0.0, 4.0\%) \\ \hline
\end{tabular}
\caption[vtp07]{Average throughput in Kbps and loss ratio for one audio and one bulk connection between nodes 0 and 7}
\label{tab:vtp07}
\end{table}

\subsection{Multiple concurrent bulk TCP transfers}

In a modified string experiment 
there is a connection between
every two neighboring nodes in a string.  In addition there is a
connection between the last and the first node that spans all nodes in
the string. This topology is used to investigate fairness between single
and multihop transfers and how hidden terminal problem affects them.

\begin{figure}
     \centering
     \subfigure[]
     {
          \label{fig:string-concurrent-norts-200}
          \includegraphics[width=.22\textwidth,height=0.4\textwidth,angle=-90]
          {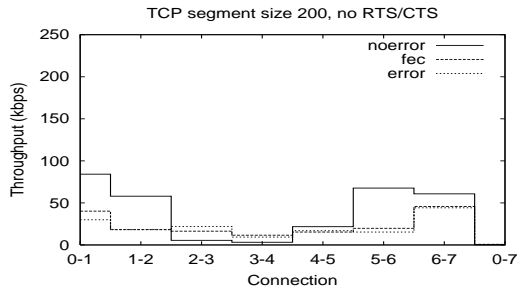}
     }
     \subfigure[]
     {
          \label{fig:string-concurrent-rts-200}
          \includegraphics[width=.22\textwidth,height=0.4\textwidth,angle=-90]
          {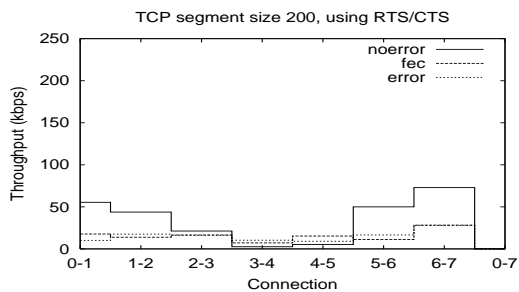}
     }
     \subfigure[]
     {
           \label{fig:string-concurrent-norts-1000}
           \includegraphics[width=.22\textwidth,height=0.4\textwidth,angle=-90]
           {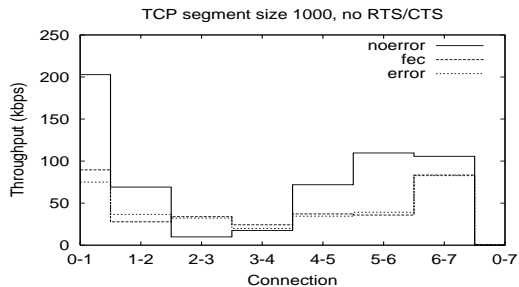}
     }
     \subfigure[]
     {
           \label{fig:string-concurrent-rts-1000}
           \includegraphics[width=.22\textwidth,height=0.4\textwidth,angle=-90]
           {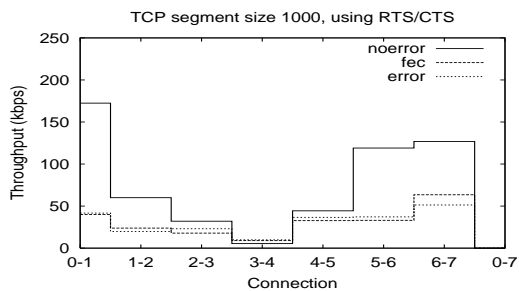}
     }
     \caption[String with concurrent transfers]{String with concurrent TCP transfers}
     \label{fig:string-concurrent}
\end{figure}

Performance of one-hop TCP connections is much better when compared
to the one multi-hop connection (figure \ref{fig:string-concurrent}).
This setup shows how a multihop connection fails in the presence of many
single hop transfers. Furthermore, since all nodes are active at the same
time, the performance of single hop connections is also affected, since
a single node cannot communicate with two different nodes simultaneously.

In the figures \ref{fig:string-concurrent-norts-200},
\ref{fig:string-concurrent-norts-1000}, where RTS/CTS is not used, it can
be shown that large packet sizes result in larger total throughput.  It is
interesting to note that FEC at 60\% efficiency is more fair to middle
connections (2-3 and 3-4) when compared to a situation either with or
without errors.  Because of errors, a single connection is never able to
fully capture the channel as is the case without errors. This results in
reduced overall throughput, and better fairness. Use of RTS/CTS (figures
\ref{fig:string-concurrent-rts-200},
and \ref{fig:string-concurrent-rts-1000}) reduces throughput and does
not improve fairness. In all cases, the single multihop connection is
able to transfer data, but at a very low rate (about 1 kbps).

\section{Mesh}
\index{Evaluation!mesh}

Mesh topology is an example of a more realistic topology than a
string from the previous section. Every node in a mesh is connected to
either two (corners), three (sides) or four (inner) other nodes. Mesh
topology allows us to see how TCP performs in a more realistic ad
environment.  There are two types of traffic passing through the mesh.
Along all the vertical paths are bulk TCP connections. For example
in a 6x6 mesh,
the nodes are numbered in a row major fashion, with the bottom row numbered
0,1,2,..5, and the top row numbered 30-35.
We established 6 TCP connections, with source nodes 30 to 35  and destination nodes 0
to 5 respectively (6 connections from the top row to the bottom row).
They are numbered 1 to 6 respectively. Along the
horizontal paths are constant bit connections.  In a 6x6 mesh, the
connections originate in nodes 0, 6, 12, 18, 24, 30 and terminate in 5,
11, 17, 23, 29, 35. (these are 6 horizontal connections between nodes
in the left-most column and the corrersponding nodes in the righ-most
column). Constant bit sources are similar to bulk sources
in that they too have unlimited supply of packets. The difference is
that constant bit (CBR) sources send packets at regular intervals.
CBR sources use TCP as transport protocol. Packet size is fixed in
all the simulations to 1000 bytes. The rate at which the CBR sources
generate packets is one of 23.3kbps or 233.3kbps. CBR sources represent
interference traffic, by introducing constant load on the network. With
23.3kbps sources, a CBR source sends one TCP segment every 1.5 seconds,
and with 233.3kbps it is 0.05 seconds. Therefore network load increases
with higher CBR source rate.

\begin{figure}
     \centering
     \subfigure[]
     {
          \label{fig:grid-200-23_3}
          \includegraphics[width=.22\textwidth,height=0.4\textwidth,angle=-90]
          {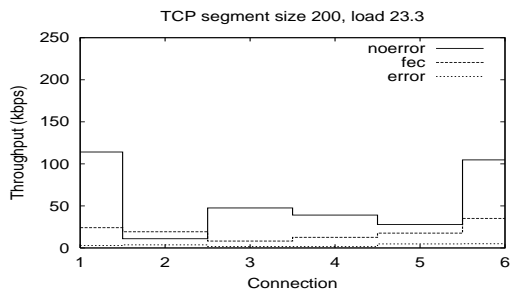}
     }
     \subfigure[]
     {
          \label{fig:grid-200-233_3}
          \includegraphics[width=.22\textwidth,height=0.4\textwidth,angle=-90]
          {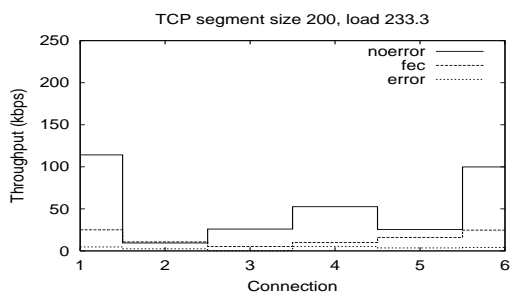}
     }
     \subfigure[]
     {
           \label{fig:grid-1000-23_3}
           \includegraphics[width=.22\textwidth,height=0.4\textwidth,angle=-90]
           {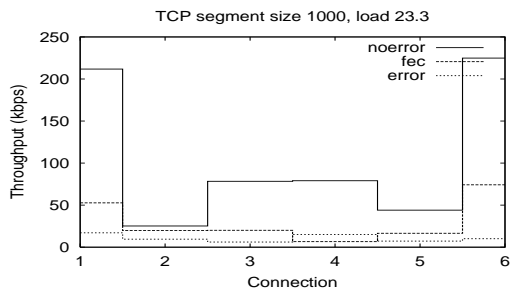}
     }
     \subfigure[]
     {
           \label{fig:grid-1000-233_3}
           \includegraphics[width=.22\textwidth,height=0.4\textwidth,angle=-90]
           {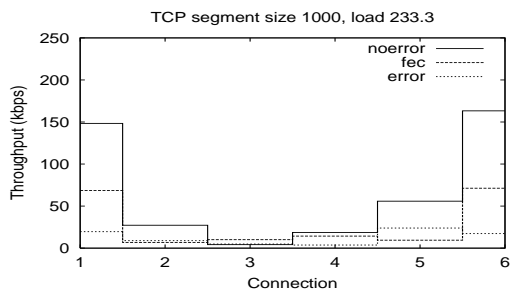}
     }
     \caption{Ad hoc mesh}
     \label{fig:mesh}
\end{figure}

Figure \ref{fig:mesh} shows results for mesh of size 6x6.  Here we
can see that in the presence of errors and without FEC, TCP is barely
able to transfer any packets, with throughput close to zero. This is
because with 6 hops, probability that a packet reaches destination
is very small. Here use of FEC helps considerably in that is allows
all connection to transfer data. Again larger packet sizes show better
throughput. As load increases, without FEC, some middle TCP connections
fail completely for small packet sizes. With FEC, throughput of bulk
connections is reasonable in the sense that it is about 20\% to 50\% on
average of the throughout without any errors. In some cases, throughput
of bulk connections with FEC is even better than their throughout would
without any errors. This is because wireless errors also affect the
throughput of interfering traffic and lower it considerably, so that
bulk TCP connections experience less interference from CBR traffic.

A similar simulation in \cite{Gerla99a} differs in that static routing
is used, wheres here DSDV is used. The consequence of this is that it is
possible for DSDV to incorrectly determine that a node is unreachable due
to wireless errors and it re-routes packets using a different route. It
is therefore possible that the bulk and CBR traffic do not flow always
along horizontal or vertical direction only.

\section{Summary and Conclusions}
In this paper we investigated the perofrmance of bulk traffic using TCP and
real-time audio traffic using UDP over an ad-hoc networ using IEEE802.11.
Our results indicates that 802.11 is suitable only for small ad-hoc networks
with number of hops 2-3. A bigger network results in a much degraded perofrmance
for both TCP and UDP.


\bibliographystyle{ieeetr}
\bibliography{ict}

\end{document}